\documentclass[conference]{IEEEtran}
\ifCLASSINFOpdf
\else
\fi

\usepackage{booktabs} 
\usepackage{cleveref}
\usepackage{framed}
\usepackage{todonotes}
\usepackage{mathptmx}
\usepackage{multirow}
\usepackage{colortbl}
\usepackage{balance}
\usepackage[scaled=.90]{helvet}
\usepackage{courier}
\usepackage[normalem]{ulem}
\usepackage{tabularx}
\usepackage{algorithm}
\usepackage{algorithmic}
\usepackage{amsmath}
\usepackage{mathtools}
\usepackage{listings}
\usepackage{url}
\usepackage{xcolor}
\usepackage{mdframed}

\newcommand{\etal}{\textit{et al.~}}

\newcommand{\CC}{\cellcolor{gray!20}}

\begin{document}
%
\title{A Simple NLP-based Approach to Support Onboarding and Retention in Open Source Communities}

 \author{\IEEEauthorblockN{Christoph Stanik, Lloyd Montgomery, Daniel Martens, Davide Fucci and Walid Maalej}
 \IEEEauthorblockA{University of Hamburg, Germany \\
 Email: \{stanik, montgomery, martens, fucci, maalej\}@informatik.uni-hamburg.de}}



%


\maketitle


\begin{abstract}
Successful open source communities are constantly looking for new members and helping them become active developers.
A common approach for developer onboarding in open source projects is to let newcomers focus on relevant yet easy-to-solve issues to familiarize themselves with the code and the community.
The goal of this research is twofold. First, we aim at automatically identifying issues that newcomers can resolve by analyzing the history of resolved issues by simply using the title and description of issues. 
Second, we aim at automatically identifying issues, that can be resolved by newcomers who later become active developers.
We mined the issue trackers of three large open source projects and extracted natural language features from the title and description of resolved issues.
In a series of experiments, we optimized and compared the accuracy of four supervised classifiers to address our research goals.
Random Forest, achieved up to 91\% precision (F1-score 72\%) towards the first goal while for the second goal, Decision Tree achieved a precision of 92\% (F1-score 91\%).
A qualitative evaluation gave insights on what information in the issue description is helpful for newcomers. 
Our approach can be used to automatically identify, label, and recommend issues for newcomers in open source software projects based only on the text of the issues.
\end{abstract}


%
\IEEEpeerreviewmaketitle

\section{Introduction}
\label{sec:introduction}

Open Source Software (OSS) projects suffer from a high turnover rate~\cite{Zhou2012}. Although expected in volunteer-based communities, this challenge needs to be addressed for a sustainable project growth. 
Retention of the \textit{newcomers}, i.e.~the community members who have made minimal (if any) contributions to the project, is thus an important objective of OSS communities.

In a systematic literature review, Steinmacher et al.~\cite{Steinmacher2015} found that one of the major barriers faced by newcomers to OSS projects is \textit{finding a way to start} contributing to the project. 
Onboarding---the stage at which an outsider makes the first contributions to the project---usually starts by picking an open issue (e.g., a bug report or a feature request) to work on from the project issue tracker.  
This can be daunting for someone without proper knowledge of the project.
Mozilla, for instance, attempts to solve the problem of ``which issues should newcomers work on'' by providing a tag, \textit{good-first-bug}, core contributors in the community attach to issues they feel are good fits for newcomers.\footnote{\url{https://wiki.mozilla.org/Good_first_bug}}
Labuschagne and Holmes found, however, that ``developers whose initial contribution is on a [good-first-bug] [...] are less likely to become long-term contributors'' \cite{Labuschagne2015}.  This imply that the use of such a tag, at least in its current form, does not solve the problem at hand.

Our research addresses the ``finding a way to start'' onboarding barrier. We use machine learning (ML) and Natural Language Processing (NLP) to identify which issues are more likely to be resolved by newcomers --- based on issue history of three large OSS projects.  
Experienced contributors to OSS projects better understand the underlying complexity of the code base when selecting and working on  issues, whereas newcomers just have what is readily available to them: the title and description of the issue. 
For this reason, we focus on configuring and training our machine learning models based on the textual description of the issues. 
During the onboarding period, newcomers already work on and resolve issues.
Therefore, training a classifier to identify which issues newcomers should resolve is our first step.
This is captured in the following research question:
\begin{description}
\item [\textbf{RQ1:}] To what extent can the issues that newcomers resolve be predicted based on their textual features?
\end{description}
The objective is to automatically identify issues (i.e., similar to what \textit{good-first-bug} does based on experts) rather than to assign issues, as it is done during issue triaging~\cite{BV13}. 

Additionally, in an effort to improve retention in OSS communities, our second objective is to predict the issues that were resolved by newcomers who later became active developers.
\begin{description}
\item[\textbf{RQ2:}] To what extent can the issues that newcomers---who subsequently become active developers---resolve be predicted based on their textual features?
\end{description}

Finally, to better understand which and to what extent textual features are helpful in selecting an issue, we first manually analyzed the content of a sample of issues used to trained the ML models, and then triangulate our results with a series of interviews.
\begin{description}
\item[\textbf{RQ3:}] What themes can be found in issues and which textual features drive newcomers and experienced contributors decision of working on an issue?
\end{description}
The main contribution of this paper is twofold.
First, we introduce a simple, NLP-based approach for automatically tagging issues as \textit{suitable for newcomers}. 
Second, we developed a framework for automatically evaluating, comparing, parameterizing, and refining supervised machine learning models for issue categorization.
Although our models were built with specific goals in mind, the framework allows for the target class and other configurations to be redefined (as we did for RQ2).
The second contribution comes as a replication package, containing the code, artifacts, and  instructions.\footnote{\url{https://mast.informatik.uni-hamburg.de/replication-packages/}}

The remaining paper is structured as follows.
Section \ref{sec:methodology} explains our methodology by defining important terms, describing our dataset, and reporting the research design.
In Section \ref{sec:evaluation} we present our results to the research questions.
Section \ref{sec:discussion} discusses the implications of our work while Section \ref{sec:limitations} elaborates on its limitations.
Section \ref{sec:related_work} presents related work by stating work regarding barriers for newcomers in OSS, retention in OSS, and machine learning approaches applied for classification problems in OSS.
In Section \ref{sec:conclusion}, we discuss future work and conclude the paper.
\section{Methodology}
\label{sec:methodology}
We first define the main concepts used in our machine learning models, and then introduce the datasets and explain the steps we followed to build the models.
Finally, we describe how we conducted the qualitative part of the study to answer RQ3.

\subsection{Definitions}
\label{ssec:definitions}
In OSS communities various member types contribute in different ways. 
According to von Krogh et al.~\cite{VonKrogh2003}, ``joiners'' are members that have not yet made a contribution; while ``newcomers'' are members that have made at least one contribution, but have not yet found a more formal role within the community. 

Nakokoji et al. \cite{Nakakoji2002} define these roles more formally as: project leader, core member, active developer, peripheral developer, bug fixer, bug reporter, reader, and passive user.
In this paper, we focus on the role of ``active developer'' as we aim at investigating issues resolved by newcomers (RQ1), who later become active developers (RQ2).
In the following we define these two relevant roles.

\textbf{Newcomer.}
A contributor to an OSS project is considered a newcomer while an initial set of issues are resolved.
Thereafter, the newcomer can transition to some other role within the organization.
Due to the ambiguity behind how long a contributor is a newcomer (and lack of published work empirically defining the term), we take into account three separate ranges of resolved issues before a contributor is no longer considered a newcomer: one, five, and ten.

Formally, let $I\big(c\big)$ be the number of issues resolved by contributor $c$, then a newcomer (i.e., $nc_t$) is defined as:
\begin{equation}
    nc_t \iff I\big(c\big) = t, t \in \{1, 5, 10\}
\end{equation}

\textbf{Active developer.}
As defined by Nakokoji et al., ``active developers regularly contribute new features and fix bugs'' \cite{Nakakoji2002}.
Later defined by Di Bella et al., active developers are contributors who ``develop limited part of the project for an extended period of time'' \cite{DiBella2013}.
These definitions do not provide a specific way to calculate which contributors are active developers.
We define a \textit{monthly active developer} as a contributor who resolves at least the project monthly median number of resolved issues.

Formally, let $I\big(c\big)_{m}$ be the issues resolved by contributor $c$ in month $m$, and $P\big(m\big)$ be the number of issues resolved in the project during the same month, then a monthly active developer (i.e., $ad\big(m\big)$) is defined as follows:
\begin{equation}
    ad\big(m\big) \iff I\big(c\big)_{m} \geq P\big(m\big)_{MED}
\end{equation}
where $P\big(m\big)_{MED}$ represents the median of $P\big(m\big)$ per contributor.
For example, for a project with $P\big(m\big)_{MED}$ = 4 in a month, a contributor $c$ with  $I\big(c\big)_{m} = 5$ for the same month is considered a monthly active developer ($ad\big(m\big)$). 

We consider $c$ an \textit{active developer} ($ad$) in the project if, at any given point in time, $c$ was a monthly active developer for six months in a row. 
We seek for contributors who ever became an active developer. It is possible for a contributor to become an active developer only for a certain period of time and then stop, e.g., when leaving the project.


\begin{table*}[ht]
\centering
\footnotesize
\caption{Characteristics of the OSS issue tracker datasets used in this study.}
\label{tab:issues}
\begin{tabular}{llrrc|rr|rr}
\toprule
                 &                       &                    &                           &                 & \multicolumn{2}{c|}{\textbf{Avg Length Title}} & \multicolumn{2}{c}{\textbf{Avg Length Description}} \\
\textbf{Project} & \textbf{Project Type} & \textbf{\# Issues} &  \textbf{\# Contrib.}     & \textbf{Period} & Characters         & Words         & Characters       & Words        \\ \midrule
\CC Qt           & \CC Cross-platform SDK& \CC 55,610         & \CC 780                   & \CC 2003 - 2017 & \CC \CC 59.38      &  \CC 8.70     &  \CC 1,127.59    &  \CC 111.89  \\
Eclipse          & IDE                   & 158,843            & 3,964                     & 2001 - 2017     &  55.58             & 7.81          & 870.23           & 81.16        \\
\CC LibreOffice  & \CC Office Suite      & \CC 10,958         & \CC 615                   & \CC 2010 - 2017 &  \CC 60.20         & \CC 9.16      & \CC 740.56       & \CC 99.75    \\ \bottomrule
\end{tabular}
\end{table*}

\subsection{Datasets}
\label{ssec:data}

We mined the issue trackers of three open source software projects, presented in \Cref{tab:issues}.
We selected Qt---a cross-platform application development framework, Eclipse---an integrated development environment (IDE), and LibreOffice---an office suite.

In total, we collected about 225,000 issues.
Comparing all three projects, the number of contributors who resolved at least two issues is the highest for Eclipse with 2,096, then 509 for Qt, and 255 for LibreOffice. 65\% of the contributors in the Qt project, 53\% in Eclipse and 41\% in the LibreOffice project resolved at least two issues.
Each issue has a field for a title and a description.

\begin{figure}[ht]
    \centering
    \includegraphics[width=\columnwidth]{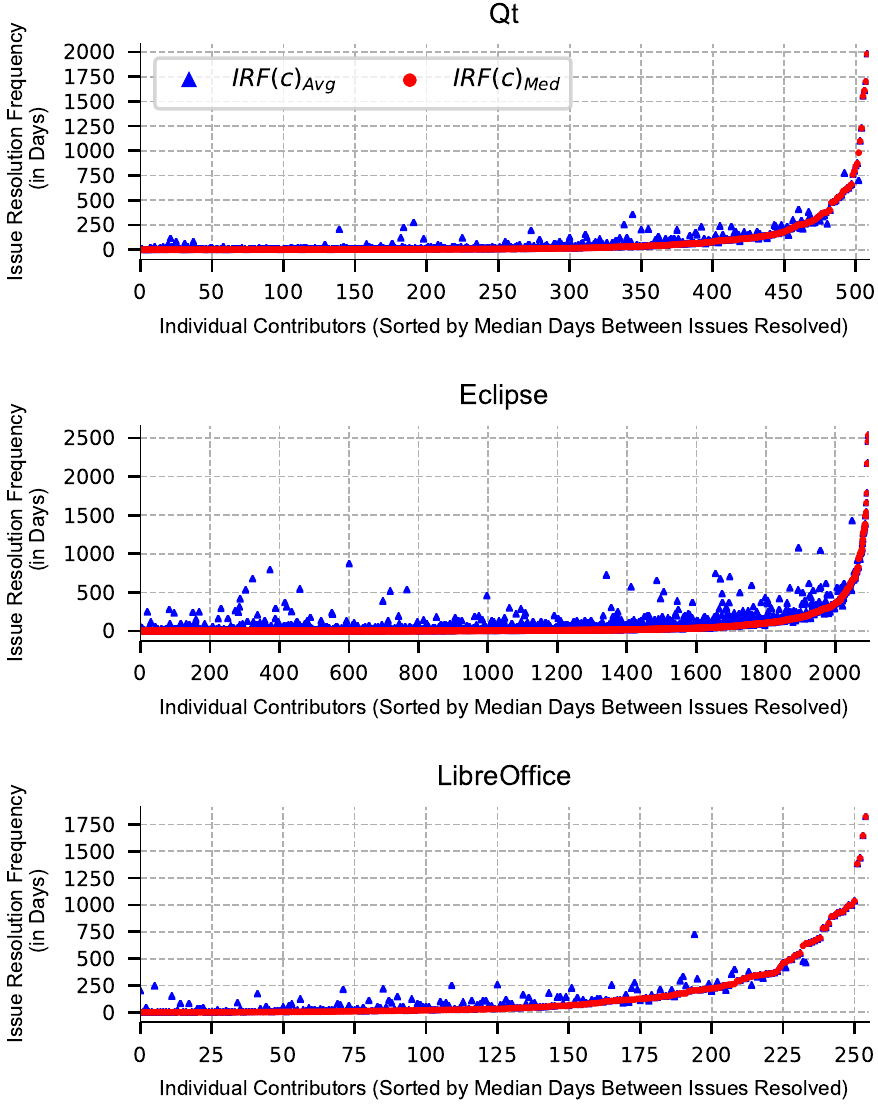}
    \caption{Average (triangles) and median (circles) Issue Resolution Frequency (IRF) for the three datasets. The x-axis reports each contributor in ascending order of IRF.}
    \label{fig:avgmeddis}
\end{figure}

Besides the textual features, we characterize the datasets in terms of contributors' issue resolution frequency (i.e., $IRF\big(c\big)$).
Figure \ref{fig:avgmeddis} reports the $IRF$ value (on the y-axis) for each contributor (on the x-axis) per project.
To account for the long-tail distribution of issues resolved over time by the project contributors, we calculated two flavors of $IRF\big(c\big)$. 
$IRF\big(c\big)_{Avg}$ is calculated by the average time passed between the issues resolved by contributor $c$, in days (triangles in \Cref{fig:avgmeddis}). 
Alternatively, $IRF\big(c\big)_{Med}$ is calculated using the median time in days (circles in \Cref{fig:avgmeddis}). Table \ref{tab:datasets-desc} summarizes the rounded $IRF$ values per project.


The goal in defining an active developer is to group contributors into an average amount of contributions, as decided by the general activity level of the project.
$P\big(m\big)_{MED}$ was selected over $P\big(m\big)_{AVG}$ because of the outliers of the long-tailed distributions in \Cref{fig:avgmeddis}.
By selecting the median over the average, the effects of the extreme values of non-active contributors (some $IRF$ values are as large as seven years) are mitigated.

\begin{figure}[b]
    \centering
    \includegraphics[width=\columnwidth]{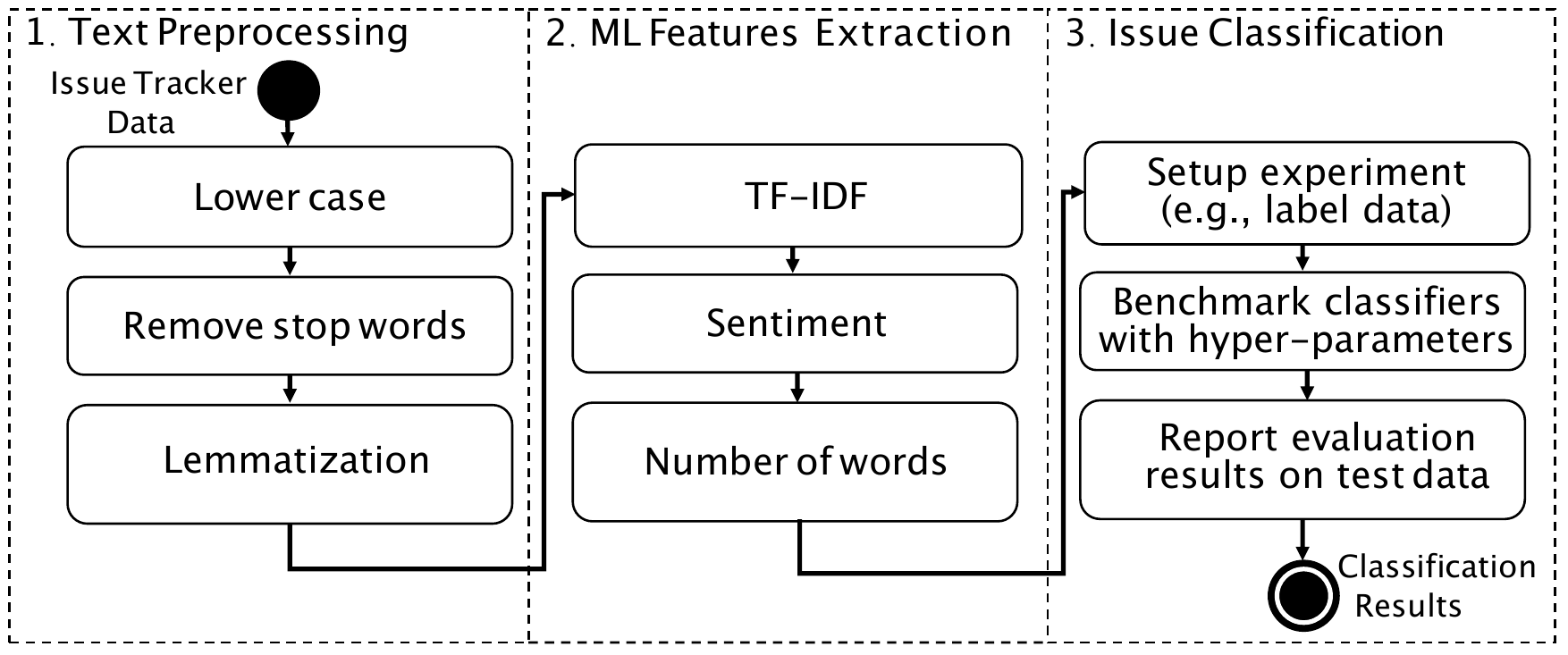}
    \caption{Steps performed for the machine learning approach.}
    \label{fig:rq1_steps}
\end{figure}

\subsection{Machine learning approach}
Figure \ref{fig:rq1_steps} summarizes the steps performed to answer the research questions.
We performed three major steps, text preprocessing, machine learning feature extraction, and classification including benchmarking.
Following previous research using NLP for issues analysis~\cite{bhattacharya2012automated, anvik2006should, anvik2011reducing, helming2010automatic, linares2012triaging}, we focus on the title and description of the issues coming from issue tracker systems.

\begin{table}[tb]
\centering
\footnotesize
\caption{Approximated statistics of contributors Issue Resolution Frequency ($IRF\big(c\big)$).}
\label{tab:datasets-desc}
\begin{tabular}{l|rrr|rrr}
\toprule
                                    & \multicolumn{3}{c|}{\textbf{IRF$\big($c$\big)_{Med}$}} & \multicolumn{3}{c}{\textbf{IRF$\big($c$\big)_{Avg}$} } \\ 
\textbf{Project}    & Avg           & Med       & SD           & Avg           & Med       & SD             \\ \midrule
\CC Qt              & \CC 83     & \CC 7  & \CC 214   & \CC 102    & \CC 28 & \CC 211     \\
Eclipse             & 66         & 5      & 198       & 103        & 33     & 211        \\
\CC LibreOffice     & \CC 165    & \CC 34 & \CC 291   & \CC 193    & \CC 87 & \CC 283     \\ \bottomrule
\end{tabular}
\end{table}

\textbf{Text preprocessing.}
First, we adapt the raw text to lower case, then filter out commonly used English stop words that do not add information to the text~\cite{bird2009natural} and finally perform lemmatization, which reduces the inflected forms of words to their root using NLTK \cite{bird2009natural} and WordNet \cite{miller1998wordnet}.
The result of this phase is a set of words which represents the input for the ML feature extraction. 

\textbf{ML features extraction.}
We extract the machine learning features needed to train a model which will perform the classifications.
As shown in Figure \ref{fig:rq1_steps}, we extract three different ML features from the issues: the term-frequency inverse document-frequency (TF-IDF), the sentiment scores, and the number of words used to describe the issue.

We extract the TF-IDF from each issue by combining all text of that particular issue (i.e., title and description) and using the elements from the resulting vector as features.
We preferred TF-IDF because it considers the weights (i.e., the importance) of a term in a given document with respect to the overall vocabulary as opposed to, for example, bag-of-words, which only describes the occurrences of specific words within documents.
Further, in existing research about closely related topics (e.g., bug triaging), TF-IDF is preferred over bag-of-words \cite{jonsson2016automated, bhattacharya2012automated, xuan2012developer}.
Second, we extract the sentiment associated with the text~\cite{thelwall2010sentiment} as we assume that sentiment can have an impact on newcomers selecting which issues to work on.
The sentiment score is calculated on the unprocessed descriptions by means of the SentiStrength \cite{thelwall2010sentiment} library, which returns two scores as humans can attach negative and positive emotions in single sentences---e.g., ``I loved this tool a lot until they turned it into worthless garbage of sofware''.
The score representing how negative a sentence is ranges from \textit{-5 extremely negative} to \textit{-1 not negative}.
Vice versa, the positive score ranges from \textit{+1 not positive} to \textit{+5 extremely positive}. 
Similarly, we consider number of words of the unprocessed descriptions as a feature for the models due to the potential impact it can have on newcomers picking issues to work on.

\textbf{Issue classification.}
We use the extracted ML features to perform classification benchmarks by comparing the results of four different supervised models: Random Forest, GaussianNB (Na\"ive Bayes), DecisionTree, and LIBSVM.
We chose these classifier implementations based on related work addressing similar problems~\cite{xuan2012developer, jonsson2016automated,bhattacharya2012automated}, and implemented them with the Scikit-learn library \cite{pedregosa2011scikit}.

To avoid over-fitting the model, we split the datasets into training and testing sets.
The test sets include a 15\% sample of the overall dataset that remained unseen in the creation and validation of different hyper-parameter combinations.
To check the performance of different hyper-parameter combinations, 10-fold cross-validation was performed by splitting the training sets into train and validation sets.
After the best hyper-parameter combination was found, the classification model was built based on this result and finally tested against the test set, which remained unseen until this step.
In this work, we report on the results of the ML algorithms applied to test sets which simulates new incoming issues to the issue tracker.

As explained in Section \ref{ssec:data}, there are nine subsets of the data created for this research (cartesian product of the three types of newcomers and the three projects).
Each of the nine subsets was unbalanced (see Table \ref{tab:data-balance}), and therefore balancing techniques had to be applied.
To address the issue of unbalanced data, we combined over-sampling and under-sampling approaches \cite{chawla2009data}. 
We extracted the test set for the classification before over-sampling was performed to avoid data being duplicated across train and test sets. 
For over-sampling the minority class, we duplicated the entries of the training set. 
We performed under-sampling for the majority class---i.e., we randomly removed instances from it so that the two classes have the same sample size. 
As a result, the classifiers were trained with a balanced set of instances.
Because under-sampling introduces the risk of removing important instances for the classification, we run the sampling strategy five times and performed \textit{Grid Search}~\cite{NIPS2011_4443} for each run.

\begin{table}[]
\centering
\footnotesize
\caption{Data balance of the classifications. Issues resolved by newcomer (\textit{New.}) vs. resolved by other (\textit{Other}). Issues resolved by newcomer who retained (\textit{Ret.}) vs. who quit (\textit{Quit}).}
\label{tab:data-balance}
\begin{tabular}{p{0.3cm}l|rc@{}r|rc@{}r|rc@{}r}
\toprule
                     &      & \multicolumn{3}{c|}{Qt} & \multicolumn{3}{c|}{Eclipse} & \multicolumn{3}{c}{LibreOffice} \\
                     &      & New. && Other & New.&& Other & New.&& Other \\ \midrule
\multirow{3}{*}{RQ1} &\CC $nc_1$  &\CC 776     &\CC   &\CC 53,714  &\CC 3,020 &\CC    &\CC 153,707   &\CC 457  &\CC   &\CC 10,438 \\
                     & $nc_5$  & 1,836   &   & 52,654  & 7,977 &    & 148,750   & 700  &   & 10,195 \\
                     &\CC $nc_{10}$ &\CC 2,941   &\CC   &\CC 51,549  &\CC 11,989&\CC    &\CC 144,738   &\CC 900  &\CC   &\CC 9,995 \\ \midrule \midrule

                     &      & Ret. && Quit& Ret.&& Quit & Ret.&& Quit \\ \cline{3-11}
RQ2                  &\CC $nc_1$  &\CC 159     &\CC   &\CC 617     &\CC 450   &\CC    &\CC 2,570     &\CC 76   &\CC   &\CC 381   \\ \bottomrule     
\end{tabular}
\end{table}

We chose \textit{Grid Search} for hyper-parameter tuning to find the optimal values for various parameters for each classifier, applied to each subset.
In contrast to Random Search---which samples hyper-parameter combinations for a fixed number of settings \cite{bergstra2012random}---\textit{Grid Search} exhaustively combines hyper-parameters of a defined grid.
For each hyper-parameter combination in the \textit{Grid Search}, we perform 10-fold cross-validation of the training set.
We optimize the hyper-parameters towards precision because for our approach it is more important to be sure about the class of a certain issue than finding all issues of a specific class. 
For example, if we have a high recall but low precision, we include false-positives that could lead to newcomers working on issues that should probably be resolved by e.g., core contributors.
Therefore, we accept that we do not find all issues newcomers should work on (low recall), but instead provide a list of issues that contain mostly newcomer-friendly issues (high precision).

\begin{table*}[tb]
\centering
\footnotesize
\caption{Results of supervised machine learning classifier benchmark for newcomers' onboarding over the three selected $nc_t$ thresholds. Highest precision reported in bold.}
\label{tab:results_rq1}
\begin{tabular}{ll|ccc|ccc|ccc}
\toprule
 &  & \multicolumn{3}{c|}{\textbf{Issue $nc_1$}} & \multicolumn{3}{c|}{\textbf{Issues $nc_5$}} & \multicolumn{3}{c}{\textbf{Issues $nc_{10}$}} \\ \cline{3-11} 
 \textbf{Project} & \textbf{Classifier}  & Precision & Recall & F1-score & Precision & Recall & F1-score & Precision & Recall & F1-score \\ \midrule

\multirow{4}{*}{Qt} &\CC Random Forest &\CC \textbf{.84} &\CC .37 &\CC .52 &\CC \textbf{.91} &\CC .59 &\CC .71 &\CC \textbf{.91} &\CC .52 &\CC .66 \\
 & GaussianNB & .38 & .56 & .46 & .43 & .65 & .52 & .46 & .73 & .57 \\ 
 &\CC DecisionTree &\CC .55 &\CC .57 &\CC .56 &\CC .79 &\CC .80 &\CC .79 &\CC .62 &\CC .63 &\CC .63 \\
 & LIBSVM & .70 & .51 & .59 & .89 & .83 & .86 & .70 & .58 & .63 \\ \midrule

\multirow{4}{*}{Eclipse} &\CC Random Forest &\CC \textbf{.91} &\CC .60 &\CC .72 &\CC  \textbf{.53} &\CC .16 &\CC .24 &\CC \textbf{.57} &\CC .18 &\CC .27 \\
 & GaussianNB & .37 & .61 & .46 & .40 & .69 & .50 & .41 & .67 & .51 \\
 &\CC DecisionTree &\CC .62 &\CC .65 &\CC .64 &\CC .46 &\CC .48 &\CC .47 &\CC .42 &\CC .45 &\CC .43 \\
 & LIBSVM & .47 & .44 & .45 & .40 & .47 & .43 & .42 & .49 & .45 \\ \midrule

\multicolumn{1}{l}{\multirow{4}{*}{LibreOffice}} &\CC Random Forest &\CC \textbf{.63} &\CC .29 &\CC .40 &\CC \textbf{.63} &\CC .25 &\CC .35 &\CC \textbf{.55} &\CC .22 &\CC .31 \\
\multicolumn{1}{l}{} & GaussianNB & .38 & .44 & .41 & .37 & .46 & .41 & .40 & .53 & .45 \\
\multicolumn{1}{l}{} &\CC DecisionTree &\CC .45 &\CC .52 &\CC .48 &\CC .43 &\CC .46 &\CC .45 &\CC .39 &\CC .42 &\CC .40 \\
\multicolumn{1}{l}{} & LIBSVM & .57 & .47 & .51  & .46 & .44 & .48 & .47 & .44 & .45 \\ 
\bottomrule
\end{tabular}
\end{table*}
\subsection{Qualitative approach}
We utilized a qualitative approach based on thematic analysis of the issues followed by a series of interviews with OSS contributors to get further insights.  

\textbf{Thematic analysis of issues}.
The goal of the thematic analysis is to get a better understanding of the issues by looking for frequent themes within the issue's title and description. The thematic analysis was conducted by the first four authors of this paper who have several years of experience in software development---a more formal description and discussion of the method can be found in Nowell et al. \cite{nowell2017thematic}.
We created a stratified random sample of 200 issues for the two models, addressing both RQ1 and RQ2.
Half of the issues belong to the positive class \textit{P} (i.e., issue resolved by a newcomer) while the other half belongs to the negative class \textit{N} (i.e., issue not resolved by a newcomer). 
We assigned an equal number of issues to be analyzed by each coder; information other than issue title and description (e.g., the class assigned to them after applying approach) was hidden to avoid bias.
For each issue, the coders looked for themes occurring in the text and noted their observations independently.
Afterwards, in an iterative fashion, the coders compared their findings and solidified them in a list of recurring themes.

\textbf{Semi-structured interviews}.     
The goals of the interview are (1) understand what role textual features play in the process of selecting issues and (2) to generate insights on how to improve our approach.  
For the kind of information sought regarding goal (1), we formulated closed questions, whereas we left participants free to expound their opinions about their process of selecting an issue to elicit insights regarding goal (2).

We purposefully sampled participants to get two homogeneous groups, newcomers to OSS projects and experienced contributors.
We were supported by community representatives of the Qt and Eclipse projects in realizing our sampling strategy.
In total, eight contributors (four from the Qt and four from the Eclipse community) agreed to be interviewed.

First, we asked contributors general questions about their experience with the project, their motivations to contribute and barriers faced as newcomers.
A second set of questions focused on the aspects considered when selecting issues to work on and, according to the participant's experience, whether such aspects are helpful for newcomers initial onboarding and retainment.
In the third part of the interview, we presented participants with issues from their respective projects and asked them to vocalize their decision-making process of selecting an issue, focusing on textual aspects of the title and description.
Afterwards, we allowed participants to give free comments and feedback\footnote{The interview script is part of the replication package}. 

Each interview was conducted by two authors over the phone or via telco software and lasted between 25 to 35 minutes. 
We recorded and transcribed the interviews, with consent from the participants, to facilitate the analysis.

For the analysis of the closed questions, we aggregated the answers from the two groups of participants. 
For the open-ended questions, we individually identified recurring themes from the interview transcripts, compared them with each other, and aggregated common ones.
\section{Results}
\label{sec:evaluation}
In this section, we report on the quantitative and qualitative results and answer our research questions. 
For the quantitative results, we show the classification benchmark across the three OSS projects and summaries of the best configuration for each classification problem. For the qualitative results, we discuss the findings from the thematic analysis of the issues and present interviews with newcomers and experts contributors.

\subsection{Identifying issues that newcomers resolve (RQ1)}
\begin{table*}[tb]
\centering
\footnotesize
\caption{Best Grid Search hyper-parameter combinations for the classifiers used to address the research questions.}
\label{tab:hyper_parameter}
\begin{tabular}{ll|c|c|c||c}
\toprule
 &  & \multicolumn{3}{c||}{\textbf{RQ1}} & \textbf{RQ2} \\
 \textbf{Project} & \textbf{Setting} & Issues $nc_1$ & Issues $nc_5$ & Issue $nc_{10}$ & Issues $nc_1$ \\
 \midrule
\multirow{2}{*}{Qt} & \CC Classifier &\CC Random Forest  &\CC Random Forest  &\CC Random Forest  &\CC Random Forest  \\
 & \vbox{\hbox{\strut Hyper-parameter}\hbox{\strut}} & 
        \vbox{\hbox{\strut max\_features = log2}\hbox{\strut n\_estimators = 3000}}   &
        \vbox{\hbox{\strut max\_features = log2}\hbox{\strut n\_estimators = 1000}} &  
        \vbox{\hbox{\strut max\_features = auto}\hbox{\strut n\_estimators = 1000}} & 
        \vbox{\hbox{\strut max\_features = log2}\hbox{\strut n\_estimators = 3000}} \\
 \midrule
\multirow{2}{*}{Eclipse} &\CC  Classifier &\CC Random Forest  &\CC Random Forest  &\CC Random Forest  &\CC Random Forest  \\
 & \vbox{\hbox{\strut Hyper-parameter}\hbox{\strut}} & 
        \vbox{\hbox{\strut max\_features = sqrt}\hbox{\strut n\_estimators = 3000}}   &
        \vbox{\hbox{\strut max\_features = log2}\hbox{\strut n\_estimators = 3000}} &  
        \vbox{\hbox{\strut max\_features = log2}\hbox{\strut n\_estimators = 3000}} & 
        \vbox{\hbox{\strut max\_features = log2}\hbox{\strut n\_estimators = 3000}} \\
 \midrule
\multirow{2}{*}{LibreOffice} &\CC  Classifier &\CC Random Forest  &\CC Random Forest  &\CC Random Forest  &\CC Decision Tree  \\
 & \vbox{\hbox{\strut Hyper-parameter}\hbox{\strut}\hbox{\strut}\hbox{\strut}} & 
        \vbox{\hbox{\strut max\_features = log2}\hbox{\strut n\_estimators = 3000}\hbox{\strut}\hbox{\strut}}   &
        \vbox{\hbox{\strut max\_features = log2}\hbox{\strut n\_estimators = 3000}\hbox{\strut}\hbox{\strut}} &  
        \vbox{\hbox{\strut max\_features = log2}\hbox{\strut n\_estimators = 3000}\hbox{\strut}\hbox{\strut}} & 
        \vbox{\hbox{\strut min samples split = 2}\hbox{\strut min samples leaf = 1}\hbox{\strut criterion = gini}\hbox{\strut splitter = best}} \\
 \bottomrule
\end{tabular}
\end{table*}
To address RQ1, we follow the steps presented in the methodology on the set of issues resolved by three types of newcomers, namely $nc_1, nc_5, nc_{10}$ (see Section~\ref{ssec:definitions}). 

Table \ref{tab:results_rq1} summarizes the results of the benchmark, presenting the classification results of the four classifiers for each project.
Table \ref{tab:hyper_parameter} shows the best working hyper-parameters for each classifier that achieves the highest precision in Table \ref{tab:results_rq1}. 
The bold values represent the best precision for each of the three sets of newcomers and each project.
We sought to automatically identify issues resolved by newcomers using binary classifiers (\textit{resolved by newcomer} and \textit{not resolved by newcomer}).
In the following, we summarize our observations for each dataset.

\textbf{Qt dataset}.
Random Forest produces the highest precision for the three sets of issues considered.
For the case of $nc_1$, it achieved a precision of 84\%---an improvement of 14\% over the second-best classifier, LIBSVM. 
When considering $nc_5$, precision raises to 91\%, and remains so for $nc_{10}$.
The model of the GaussianNB has the lowest precision in the considered dataset with a score of 38\% but a  recall 19\% higher than Random Forest.
When comparing the results between the three sets $nc_1$, $nc_5$, $nc_{10}$, the performance of the classifiers increases from $nc_1$ and $nc_5$, but stays the same when observing $nc_5$ to $nc_{10}$.

\textbf{Eclipse dataset}.
In the Eclipse dataset, Random Forest attained precision of 91\% for $nc_1$, with a recall of 60\%, resulting in an F1-score of 72\%. 
The hyper-parameters used to achieve this result are \textit{n\_estimators = 3000} and \textit{max\_features = sqrt} (see Table \ref{tab:hyper_parameter}).
The second best classifier is the Decision Tree with a precision of 62\%---a loss of 29\% when compared to Random Forest.
Again, the GaussianNB classifier achieved the lowest precision but a comparative recall when compared to the other models.
The precision of this model decreases by 54\% compared to Random Forest.
In the transition between the three sets of newcomers, we observe that the precision for Random Forest decreases by 38\% for $nc_5$, and 34\% for $nc_{10}$.
Table \ref{tab:results_rq1} shows that the GaussianNB classifier, although producing rather low scores, yields similar results for all sets of issues ($nc_1$, $nc_5$, and $nc_{10}$). 

\textbf{LibreOffice dataset}.
Similar to the Qt dataset, Random Forest outperforms the other classifiers when optimizing for precision over the three sets of issues.
Random Forest achieved a precision of 63\% for $nc_1$ and $nc_5$, which decreases to 55\% for $nc_{10}$. 
The Decision Tree and GaussianNB classifiers both have a comparably low precision below 50\% for all sets of issues.
In contrast to the best model, the GaussianNB's precision is 25\% lower.
As it happens for the other datasets, increasing the number of issues, especially from $nc_5$ to $nc_{10}$, does not seem to improve recall and F1 (except for the Na\"ive Bayse case).
In contrast to the other two projects, the precision and recall for the LibreOffice dataset are, in general, lower.

\textbf{Summary.}
In general, Random Forest outperforms the other classifiers, regardless of the project and the range of issues considered, while the GaussianNB model underperformed in all cases.
Table \ref{tab:hyper_parameter} shows the best working classifier and hyper-parameter combinations for all sets of issues.
The table reveals, that in most cases, the best working hyper-parameters for Random Forest are \textit{n\_estimators = 3000} and \textit{max\_features = log2}.
Since \textit{n\_estimators = 3000} is the outer boundary in our \textit{Grid Search} grid, we assume that we might get better classification results when we search close to the value of 3000 or when we extend the outer boundary.
The grid of hyper-parameters used for the classifiers can be found in the replication package.

\begin{mdframed}
\textbf{Answer to RQ1:} We are able to identify the issues newcomers can resolve using only textual features, with an average precision (across all three projects) of 79\%, 69\%, and 68\% for $nc_1$, $nc_5$, and $nc_{10}$, respectively.
\end{mdframed}

\subsection{Identifying issues that newcomers---who later become active developers---will resolve (RQ2)}
\begin{table}[tb]
\footnotesize
\caption{Results of supervised machine learning classifier benchmark for newcomers' retention. Highest precision reported in bold.}
\label{tab:results_rq2}
\begin{tabular}{llccc}
\toprule
\textbf{Project} & \textbf{Classifier}    & \textbf{Precision} & \textbf{Recall} & \textbf{F1-score} \\ \midrule
\multirow{4}{*}{Qt} 
                    & \CC Random Forest  & \CC \textbf{.81}  & \CC .59  & \CC .67 \\
                    & GaussianNB         &     .68  &     .35  &     .45 \\
                    & \CC DecisionTree   & \CC .73  & \CC .78  & \CC .75 \\
                    & LIBSVM             &     .38  &     .71  &     .49 \\
\midrule
\multirow{4}{*}{Eclipse} 
                    & \CC Random Forest  & \CC \textbf{.67}  & \CC .41  & \CC .50 \\
                    & GaussianNB         &     .62  &     .56  &     .59 \\
                    & \CC DecisionTree   & \CC .59  & \CC .66  & \CC .62 \\
                    & LIBSVM             &     .41  &     .34  &     .35 \\
\midrule
\multirow{4}{*}{LibreOffice} 
                    & \CC Random Forest  & \CC .88  & \CC .88   & \CC .88 \\
                    & GaussianNB         &     .82  &     .28   &     .41 \\
                    & \CC DecisionTree   & \CC \textbf{.92}     & \CC .92  & \CC .91 \\
                    & LIBSVM             &     .29  &     .40  &     .33  \\
\bottomrule
\end{tabular}
\end{table}
To address this research question, we only consider issues resolved by $nc_1$.
This decision is due to the time complexity of the algorithms, and justified by $nc_1$ having the highest average precision in the models built to answer RQ1.
Using this set, we automatically labeled each issue as \textit{resolved by active developer} or \textit{not resolved by active developer} following the definition in \Cref{ssec:definitions}.
We then trained the four binary classifiers over this new dataset.

Table \ref{tab:results_rq2} summarizes the results of the classification benchmark used to answer RQ2.
We summarized our observations for each dataset.
The highest precision of 92\% was achieved by the Decision Tree classifier in the LibreOffice project. 
The best hyper-parameter combinations regarding this RQ are addressed in Table \ref{tab:hyper_parameter}.

\textbf{Qt dataset.}
In the Qt dataset, the highest precision was achieved by Random Forest at 81\%.
The hyper-parameters that achieved this result in the Qt project are: \textit{n\_estimators~=~3000} and \textit{max\_features~=~log2}.
The second best classifier in our experiment is the Decision Tree which has balanced values of precision and recall, which are above 70\%.
Even though LIBSVM achieved a comparably high recall of 71\%, its precision is behind the results of Random Forest by 43\%.

\textbf{Eclipse dataset.}
The highest precision achieved is 67\% using Random Forest. 
The best combination of hyper-parameters for Random Forest are \textit{n\_estimators = 1000} and \textit{max\_features = log}.
The GaussianNB classifier achieved a similar precision (5\% less) as the Random Forest classifier but was able to attain a 15\% higher recall.
The Decision Tree achieved the highest recall of 78\%, but only a precision of 59\%, which makes it the better choice if the goal is to find all issues of interest and false positives are not a problem.
Again, LIBSVM achieved the lowest score of the four classification models, 18\% less precision with respect to the next best model.

\textbf{LibreOffice dataset.}
The hyper-parameters of the Decision Tree that achieve the best result are \textit{criterion = gini},  \textit{splitter = best}, \textit{min\_samples\_split = 2}, and \textit{min\_samples\_leaf = 1}.
The Decision Tree is closely followed by the Random Forest classifier, which achieved 88\% precision and recall by means of the hyper-parameters: \textit{n\_estimators = 100} and \textit{max\_features = auto}.
As for the other datasets, LIBSVM performed worse with a decrease of 63\% when compared to the results of the Decision Tree.

\textbf{Summary.} The classification for the Qt and the LibreOffice dataset show promising results with a precision of at least 81\%.
In the Eclipse dataset, the classifiers achieved similar results, but Random Forest still outperforms the other classifiers. 
We observe that the best working hyper-parameters are similar to the one reported for RQ1 (see Table \ref{tab:hyper_parameter}).
LIBSVM might not be suitable to address RQ2 as it achieved the worst results. 

\begin{mdframed}
\textbf{Answer to RQ2:} We are able to identify issues that newcomers---who later become active developers---can resolve, using only textual features, with an average precision (across all three projects) of 79\%.
\end{mdframed}

\subsection{Qualitative evaluation to generate insights (RQ3)}
\begin{table}[tb]
\centering
\footnotesize
\caption{Results of the content analysis showing the theme and their counts for both models. 
    \textit{RQ1 \{P$=$issue resolved by $nc_1$ newcomer.} \textit{N$=$issue resolved by contributor.\}}, \textit{RQ2 \{P$=$issue resolved by $nc_1$ newcomer who retained in the project.} \textit{N$=$issue resolved by $nc_1$ newcomer that quit the project.\}}
    }
\label{tab:content_analysis}
\begin{tabular}{clc|c||c|c}
\toprule
 &        & \multicolumn{2}{c||}{\textbf{RQ1 Issues}}   & \multicolumn{2}{c}{\textbf{RQ2 Issues}}               \\
 &                              & \multicolumn{2}{c||}{Count}   &  \multicolumn{2}{c}{Count}    \\ 
\#      & Theme                             & P     & N             & P     & N                 \\
\midrule
\CC T1  & \CC Easy fix                      &\CC 22 &\CC 25         &\CC 26 &\CC 12             \\
    T2  &     Includes steps to reproduce   &    10 &     6         &    7  &     8             \\
\CC T3  & \CC Discusses design decisions    &\CC 4  &\CC  9         &\CC 2  &\CC  9             \\
    T4  &     Complex issue                 &    6  &     7         &    6  &     5             \\
\CC T5  & \CC Includes possible fix         &\CC 4  &\CC  5         &\CC 2  &\CC  5             \\
    T6  &     Includes code examples        &    5  &     6         &    0  &     0             \\
\CC T7  & \CC Includes stack trace          &\CC 2  &\CC  0         &\CC 4  &\CC  3             \\
    T8  &     Clear and short description   &    5  &     2         &    4  &     2             \\
\CC T9  & \CC No need to change code        &\CC 2  &\CC  0         &\CC 0  &\CC  0             \\
\bottomrule
\end{tabular}
\end{table}

\textbf{Thematic analysis of issues.}
Here, we report the thematic analysis of some issues in the sample used to build the ML.
From the model used to answer \textbf{RQ1}, we consider issues of $nc_1$ as we got the best classification results for this type of newcomers. 
The result of the thematic analysis is summarized in Table \ref{tab:content_analysis}.
Based on the size of the projects, we randomly selected a stratified sample of 200 issues---4.91\% from LibreOffice, 24.53\% from Qt, and 70.56\% from the Eclipse project.

The thematic analysis of the issues revealed that newcomers work on issues that contain a stack trace, include steps to reproduce, have a clear and short description, and do not include code changes like updates of the documentation. 
As an example, the following issue was resolved by a $nc_1$ newcomer---it only asks for updating the documentation.
\begin{mdframed}[backgroundcolor=gray!20] 
\textbf{Title}:
Remove ATF references from Docs
\\\textbf{Description}:
Remove any references to ATF from docs.
\end{mdframed}
We found that issues which can be considered as easy to fix (e.g., change the label of a button) are equally resolved by $nc_1$ newcomers and more experienced contributors. 
In addition to that, issues that already include a possible fix in the description are also equally resolved.
This could be because $nc_1$ newcomers might work on several easy to fix issues but for $nc_1$ we do not consider them as newcomers after resolving the first issue and, additionally, more experienced contributors can resolve easy to fix issues, too.
Experienced contributors tend to resolve issues that discuss design decisions, like in the following example.
\begin{mdframed}[backgroundcolor=gray!20] 
\textbf{Title}:
Remove ``to'' version parameter from QML\_DEFINE\_TYPE() macros
\\\textbf{Description}:
The time has come for this to go :) If we decide we want a more flexible versioning system in the future, we can re-add it then.
\end{mdframed}

From the model of \textbf{RQ2}, the proportions of the strata are different, as we consider only the first issue each assignee worked on. The percentage share of the 200 randomly selected issues is 10.75\% for LibreOffice,  18.25\% for Qt, and 71.01\% for Eclipse.

In contrast to our findings regarding RQ1, newcomers retained in the project are more likely to work on issues we consider as a rather easy fix (26 of 38 easy to fix issues were resolved by newcomers that were retained). 
In our sample, there were no additional hints in the issues' text to distinguish between issues resolved by newcomers who retained and those who did not---e.g., issues including stack traces and steps to reproduce were equally addressed by retained newcomers and by those that quit the project.
Similarly to RQ1, also in this sample we observed that newcomers retained in the project are less likely to pick an issue that discusses a design decision.

\textbf{Interviews with OSS contributors.}
\begin{table}[!t]
    \centering
    \footnotesize
    \caption{Overview of the interview participants.}
    \begin{tabular}{llll}
        \toprule
        \textbf{\#} & \textbf{Project}  & \textbf{Resolved Issues} & \textbf{Experience}   \\ 
        \midrule
        \CC P1      & \CC   Eclipse     & \CC   40                  & \CC   2 years         \\ 
            P2      &       Eclipse     &       20                  &       3 months        \\ 
        \CC P3      & \CC   Eclipse     & \CC   $>$150              & \CC   1.5 years       \\ 
            P4      &       Eclipse     &       $>$150              &        6 years        \\ 
        \CC P5      & \CC   Qt          & \CC   $>$150              & \CC   5 years         \\ 
            P6      &       Qt          &       50                  &       1.5 years       \\ 
        \CC P7      & \CC   Qt          & \CC   $>$150              & \CC    2 years        \\ 
            P8      &       Qt          &       10-100              &        2.5 years      \\ 
        \bottomrule
  \end{tabular}
  \label{tab:participants}
\end{table}
To further understand issues that are resolved by newcomers and the role they play in their retainment, we contacted eight OSS contributors to conduct semi-structured interviews---four from the Qt and four from the Eclipse project. 
Table \ref{tab:participants} summarizes the the participants of the interviews.
First, we asked general questions about their motivations to contribute and barriers faced as newcomers.
P1, P2, P7, P8 state that part of their motivation to contribute in an OSS project is that they can have an impact on the community by improving the software.
P1, P3, P6 explicitly say that they already had experience with that software, which motivated them to contribute, too.
Typical barriers for newcomers are the maturity of the project (P1, P2) and the difficulty to find supporting material, such as documentation on best practices (P1, P2, P8).
Other barriers are, for example, the contributor's personal reputation within the community (P7), the usability of the standard tools used by the community, the difficulty of setting up the development environment (P1, P3, P5), and hard to locate code within the project code base (P2).
The answers confirm the findings in Steinmacher et al. \cite{Steinmacher2015} who show that these barriers are identified in the literature.

Subsequently, we asked questions focusing on the process of selecting issues from the project tracker and on aspects that are important to newcomers.
P1, P2, P3, and P8 find helpful the presence of a mentor who supports their decision when selecting proper issues for them, however, from the data we collected, it appears that not all contributors are mentored.
According to the participants, a mentor is an experienced contributor that is knowledgeable about the code base and knows the steps to resolve certain issues.
Research supports the necessity and the positive influence of mentors for newcomers to OSS projects~\cite{CDO12, SWG12}.

When considering the title and description, the text should be consistent between the two (P2), be specific and clear (P1, P2, P3), contain tags which makes the title easy to search (P3, P8), contain logs and stacktraces (P5, P7), and contain steps to reproduce (P1, P5, P7, P8).
Another method for selecting issues are, for example, the issue priority (P1, P5, P6) and personal interest (P3, P6, P7).
The latter are sometimes stated as \textit{tags} in the title of an issue.
These answers support the findings of the thematic analysis, as well as findings in the literature \cite{bettenburg:2007eclipse, zimmermann2010goodbugreport, schroter2010stacktraces} which showed that issues containing a stack trace are considered helpful.
In future research, these items can be considered as features for improving our classifiers.
\begin{mdframed}
\textbf{Answer to RQ3:} Newcomers find textual features such as stack traces and steps to reproduce helpful. Experienced contributors confirm that these features are important as they help to locate the issue within a large code base. The thematic analysis revealed that newcomers choose issues containing such textual information.
\end{mdframed}
\section{Discussion}
\label{sec:discussion}
This section discusses the implications of our findings for practitioners and researchers.

\textbf{Recommending issues for contributor onboarding.} 
Our results show that Random Forest is the best model to classify, for all datasets, issues that are likely to be resolved by newcomers.
The classifier performance increases from $nc_1$ to $nc_5$.
For Qt and LibreOffice dataset, the text of $nc_5$ issues contains enough information to assess future issues which will also be resolved by a newcomer, whereas considering only the first one the best precision is yielded for the Eclipse dataset.
Moreover, classifications based on $nc_{10}$ issues did not improve across the datasets.

The implication is that, using only the textual description, a recommender system can support the triaging team in deciding whether to assign an issue to a newcomer. 
In particular, when manual triaging is not applied to all the issues submitted in the tracker, our model can be used (e.g., through an ad-hoc plugin targeting the issue tracking system) to automatically tag issues that a newcomer can likely resolve.
Similarly, commercial software companies that use issue trackers can be interested in this approach to facilitate the onboarding of new hires.
Additionally, our approach can be used to support mentors in OSS projects to select newcomer friendly issues either by recommending concrete issues or by filtering them (switch the model to focus on recall).

\textbf{Recommending issues for contributor retention.} 
Random Forest is the best model to classify issues that were resolved by newcomers who were retained in the project for the Qt and Eclipse dataset, whereas Decision Tree shows better performance in the case of LibreOffice.  
In particular, for Qt and LibreOffice, the issues identified through the second model we propose can support newcomers' career advancement. 
From the newcomer perspective, it is desirable to have a mechanism that helps them improve their status in the community as it can bring professional career advantages in contexts other than OSS~\cite{HNH03}. 
From the perspective of community management, the ability to identify and recommend these issues can support the retention of newcomers and lessen turnover---particularly for the parts of the system where more contributions are needed.

\textbf{Implications of the study design}. 
When building the classifiers, we favor precision because---from an application perspective (e.g., recommender systems)---it is more important to retrieve newcomer-friendly issues at the cost of not identifying them all.
Conversely, we minimize false positives (i.e., issues that in reality are not suitable) as this can impact newcomers motivation to further engage with the community. 




\textbf{Implications for further research on issue trackers}.
We provide a classifier with automated parametrization and target class refinement which uses \textit{Grid Search} to evaluate given hyper-parameter combinations.
Although the set of parameters yields a configuration that can be already applied in practice (e.g., for the recommender systems scenario), researchers interested in replicating or extending this study can modify hyper-parameter combinations or use other techniques like \textit{Random Search}\cite{NIPS2011_4443, bergstra2012random}.
This can lead to an increased recall, which is particularly interesting from an application perspective.
For example, when creating guidelines for writing ``newcomer-friendly'' issues, human experts need all the available evidence (recall), but they can manually remove or ignore noise.

One field for further exploration is the applicability of the models across multiple projects---can one use the model trained on the Eclipse dataset to identify issues newcomers can resolve in other open-source projects?

\section{Limitations}
\label{sec:limitations}
When interpreting the results and their implications, however, the following threats should be taken into account.

\textbf{Construct validity.}
The definitions of \textit{newcomer} and \textit{active developer} (see~\Cref{ssec:definitions}) are based on the literature.
These definitions, however, are not specific enough to operationalize them for the purposes of this research.
Therefore, we pieced them together with our understanding of the OSS community to create the constructs used in this study.
We recognize the threat to validity in self-assigning such constructs.
To mitigate this threat, we follow the previous definitions as closely as we could, and used data-driven results from the projects in our dataset to shape the final assignment of the terms.
For example, in assigning the label ``active developer'' to contributors, we used medians of the datasets as the comparison point for consideration of the label, instead of assigning an arbitrary number such as ``seven issues a month.''
Moreover, to mitigate the threat of measuring the newcomer construct, we use three different thresholds. 
The decision to ground the assignment of these labels in data-driven results helps mitigate this threat.

\textbf{External validity.}
Although our calculations are self-adjusting to the size of the project (including the number of issues and contributors), this research deals only with three specific OSS projects.
However, the concepts being addressed in this research (i.e., the different roles, the use of issues trackers) apply to most, if not all, OSS projects.
To mitigate this threat, the OSS projects selected are diverse across multiple dimensions, issue tracker systems used, number of contributors, number of issues, median time between issue resolution, and product domain.

The projects follow different practices (e.g., for bug triaging), have different backgrounds, resources, and goals. 
These properties can affect the generalizability of the results as they are not taken into account when building the models.
For example, developer reputation has an impact on issue assignment and resolution in specific scenarios~\cite{BN11}. 

We addressed the representativeness of the interview participants by directly asking from the project community managers to suggest us  contributors suitable for our study. 
During the interviews, we asked specific questions about the participants' experience and made sure to have included newcomers and experienced contributors.
However, one limitation of our study is the lack of participants from LibreOffice community, as we could not reach for contributors to this project.
Therefore, the qualitative results might not apply to that community.

\textbf{Internal validity.} 
This study does not aim to identify a direct causal relationship between issues solved by newcomers and the success of their onboarding or retainment in the community. 
For RQ1 this would imply, that the only factors newcomers consider when working on issues they eventually resolve are the title and description, when in fact other factors, such as communication (e.g., on the project mailing lists), member suggestions, personal interest can also play a role in this decision.
For RQ2, causality would be extended also to imply that the issue itself has a direct relationship towards the success of newcomers in OSS projects.
We make no such claim in this research. 
However, we show that there is a relationship between the issues resolved by newcomers and their success within our dataset which can be used to make accurate prediction.

We gathered data only from the project issue tracker systems; nevertheless, there can be other channels which the community uses for submitting, assigning or selecting issues.
However, according to the guidelines for issue reporting of the three projects, issue trackers are the recommended channel. 
We acknowledge that not all the issues are publicly available on the project issue tracker---e.g., in case of security-related issues. 
However, in general, these represent only a small set compared to the rest~\cite{GBL13}.
Similarly, we do not distinguish between different types of issues (e.g., bugs vs. feature requests), although this can affect which issue(s) a newcomer tackles when joining the project.

Regarding the qualitative analysis, we strengthen internal validity having more than one coder reviewing the themes emerged from the content analysis as well as from the open-ended questions.
We mitigated possible researcher bias in the interview script (e.g., asking leading questions) by running a pilot with one of the author.  
Moreover, recording the interviews allowed us to make sure that the observations documented are those of the participants and not a shorter form recorded by the researchers. 
\section{Related Work}
\label{sec:related_work}
In this section, we investigate onboarding and retention of newcomers in OSS projects using issue tracker data.

\subsection{Newcomers onboarding in OSS}
OSS projects rely on the contributions of volunteers to build their software, however, newcomers to these communities often face barriers when trying to contribute \cite{Dagenais2010}.
Steinmacher et al. conducted a systematic literature review showing that ``social interaction,'' ``newcomers' previous knowledge,'' ``technical hurdles,'' ``documentation,'' and ``finding a way to start'' are the five main barriers for newcomers contributions~\cite{Steinmacher2015}.

For example, \textit{finding a way to start} is a barrier that newcomers experience when deciding where to first contribute in an OSS project---they adopt several strategies to address this barrier.
A large case study reports that some of them advertise their skills on the community communication channels (e.g., forums)~\cite{VonKrogh2003}, in other cases the community itself offers support by establishing mentoring programs~\cite{Labuschagne2015, CDO12}.
Finally, a newcomer can decide to pick a task to work on from the issue tracker system. 
The results of our study can support newcomers in deciding the issue to work on next so that they are likely to close it, given their recent history.

The Mozilla Foundation supports newcomers looking for issues to tackle by letting experienced contributors tag ``newcomers-friendly'' issues as \textit{good-first-bug} (GFB).
However, a quantitative study of Mozilla Core, Firefox, and Firefox OS shows that GFB bugs are less likely to be resolved than others (67\% vs. 73\% success rate)~\cite{Labuschagne2015}.
GFB starters are 63\% more likely to make at least two contributions, though they are less likely to keep contributing than non-GFB~\cite{Labuschagne2015}.

Other approaches, aimed to help newcomers finding suitable issues, leverage visual representations.
The one presented by Wang and Sarma~\cite{WS11} is based on an extension of Tesseract---an interactive visual exploration tool of OSS projects which uses metadata collected from communication channels, code repository, and issue tracker~\cite{SMW09}.  
In this context, a controlled experiment involving 27 graduate students working on four OSS projects showed that visualization tools can provide valuable information for supporting the choice that newcomers make~\cite{YJ09}. 
Another approach is to suggest mentors to newcomers. 
The mentors, expert members of the community, will then help the new members to find a suitable issue to contribute to the community~\cite{SWG12}.
Canfora \etal~\cite{CDO12} built a mentor recommender system---achieving a precision between 65\% and 71\%---\textit{YODA} using data mined from five large OSS projects communication channels and source code versioning systems.
In contrast to these  approaches, we use only data gathered from issue trackers. 

\subsection{Newcomers retention in OSS}
Alongside the issue of finding a task to work on, the literature focuses on the barriers to newcomers retention and their transition to core contributors. 
Zhou and Mockus \cite{Zhou2012} mined issue trackers data to measure contribution participation in the Gnome and Mozilla projects.
They model a contributor's willingness and opportunities---e.g., the project popularity and sociality level--- and show that the probability of a newcomer becoming a core contributor is associated with the contributor's willingness.
Initially contributing with a comment on an existing issue doubles the odds of becoming a core contributor.

Jensen \etal\cite{Jensen2011} used the interaction between the newcomers and the community over mailing lists, for four large OSS projects. 
The vast majority of newcomers (~77\%) receive an answer about a problem they need solving; however, this does not assure long-term commitment. 
In fact, only 1 in 16.67 newcomers decide to stay, a rate which is not sustainable for most OSS project.
A similar study of the Hadoop community mailing list and issue tracker \cite{SWC12} shows that less than 20\% of newcomers became core contributors. 
These investigations leveraged the social interactions (i.e., a social graph) among contributors, but not the actual text used in the communication.
Our work shows that natural text features of issues can be considered as a factor when investigating the transition of community members from newcomers to more active roles.

\subsection{Using NLP to analyze issue tracker data}
Our results can be exploited by the OSS community (e.g., developers responsible for a component, decision-makers in the project steering committee) to support bug triaging, especially when deciding whether a bug is suitable for a newcomer. 
Previous research concentrates on ways of automating the triaging process using NLP-based features \cite{wang2014fixercache, alenezi2013efficient, murphy2004automatic}.

By leveraging large-scale proprietary industry data, Jonsson et al.~\cite{jonsson2016automated} reported an approach for assigning a bug to a developer who will most likely fix it.
As in our study, they used TF-IDF to represent features from the title and description of the issue.
They employed Stacked Generalization that combines the results of multiple machine learning classifiers and achieved an accuracy between 50\% and 89\%.
This show that the ensemble approach outperforms individual classifiers.

Badashian et al.~\cite{badashian2015crowdsourced} combine NLP features extracted from \textit{Stack Overflow} tags and issue keywords to identify the expertise of a developer. 
The authors used more than 7000 issues mined from the 20 most active GitHub projects and achieved a Top-1 accuracy of 45\% and a Top-5 accuracy of 89\%. 

Finally, Mani et al.~\cite{mani2018deeptriage} use NLP features together with an attention-based deep bidirectional recurrent neural network (DBRNN-A) to tackle the problem of bug triaging. 
In particular, their model overcomes the limitation of text-based feature models, such as bag-of-words, that do not consider the syntactical and sequence information of the unstructured text. 
As in our study, the authors chose the title and the description of the issues as the input for the classification. 
Using more than 850,000 issues from Chromium, Mozilla Core, and Mozilla Firefox, they  show that DBRNN-A achieves an improvement of 12-15\% on Rank-10 and a general performance ranging between 37-43\% when compared to other classifiers.
\section{Conclusion and Future Work}
\label{sec:conclusion}
In this work, we identify issues that newcomers in OSS projects can successfully resolve. 
Moreover, we automatically identify issues resolved by newcomers who became active developers.
To that end, we first defined newcomers and active developers analytically based on existing research.
Then, we crawled data of three OSS projects (i.e., Qt, Eclipse, and LibreOffice) and developed four supervised machine learning classifiers, using simple NLP features (i.e., the issue title, description, and sentiment).
Our results, after hyper-parameters tuning, show that we are able to identify issues newcomers resolved in a project (RQ1) with high precision.
Further, we show that we are able to identify issues that newcomers---who will later become active developers---resolved (RQ2).
Other classifiers reported in the related work (e.g., REPTree and Nearest Neighbor) and additional NLP features (e.g., semantic vectors, word embeddings, bag-of-words) may be assessed in future work.
Finally, we performed a qualitative evaluation that generated insights on what textual features newcomers and contributors consider when selecting issues.
These insights can be used to further improve the classification results.
Moreover, existing research gives a qualitative definition of OSS roles (on which we based our work).
To the best of our knowledge, no generalized, analytical definition exists which can support researchers in further empirical work about role transition in OSS communities.
In this work, we introduced such definition and plan to extend it through qualitative investigations.

\section*{acknowledgement}
This work is partly funded by the H2020
EU research project OPENREQ (ID 732463).



%
\bibliographystyle{IEEEtran}
\bibliography{main.bib} 

\end{document}